\documentclass[12pt]{article}
\usepackage{graphics}
\thicklines
\newcommand{\EQ}{\begin{equation}}
\newcommand{\EN}{\end{equation}}
\newcommand{\bea}{\begin{eqnarray}}
\newcommand{\ena}{\end{eqnarray}}
\newcommand{\eea}{\end{eqnarray}}

\def\del{\Delta}
\def\ddel{{}^\bullet\! \Delta}
\def\deld{\Delta^{\hskip -.5mm \bullet}}
\def\dddel{{}^{\bullet \bullet} \! \Delta}
\def\ddeld{{}^{\bullet}\! \Delta^{\hskip -.5mm \bullet}}

\def\la{\langle}
\def\ra{\rangle}

\begin{document}
%--------------------------------------------------------------------+
\begin{flushright}
\begin{minipage}{0.25\textwidth} hep-th/0007105 \\
YITP-00-17
\end{minipage}
\end{flushright}
%--------------------------------------------------------------------+
\begin{center}
\bigskip\bigskip\bigskip
{\bf\large{Dimensional regularization of the path integral 
in curved space on an infinite time interval}}
\vskip 1cm
\bigskip
F. Bastianelli $^a$\footnote{E-mail: bastianelli@bo.infn.it}, 
O. Corradini $^b$\footnote{E-mail: olindo@insti.physics.sunysb.edu} and 
P. van Nieuwenhuizen $^b$\footnote{E-mail: vannieu@insti.physics.sunysb.edu}  
\\[.4cm]
{\em $^a$ Dipartimento  di Fisica, Universit\`a di Bologna \\ and \\
 INFN, Sezione di Bologna\\ 
via Irnerio 46, I-40126 Bologna, Italy} \\[.4cm]
{\em $^b$ C. N. Yang Institute for Theoretical Physics \\
State University of New York at Stony Brook \\
Stony Brook, New York, 11794-3840, USA}\\
\end{center}
\baselineskip=18pt
\vskip 2.3cm

\centerline{\large{\bf Abstract}}
\vspace{.8cm}

We use dimensional regularization 
to evaluate quantum mechanical path integrals in arbitrary curved spaces
on an infinite time interval. We perform 3-loop calculations in Riemann 
normal coordinates, and 2-loop calculations in general coordinates.
It is shown that one only needs a covariant two-loop counterterm 
($V_{DR}={\hbar^2\over 8} R$)
to obtain the same results as obtained earlier in other regularization 
schemes.  
It is also shown that the mass term needed in order to avoid infrared 
divergences explicitly breaks general covariance in the final result.

\newpage

%====================================================================+
\section{Introduction}

The path integral formulation of quantum mechanics \cite{Feyn} 
is quite subtle when applied to particles moving in a curved 
space \cite{somerefs}.
It can be used to evaluate anomalies in quantum field theories, but 
only when the corresponding quantum mechanical models are defined on a 
finite interval of the worldline.
When viewed as one dimensional QFTs on the worldline (but with
higher-dimensional target spaces), one is dealing with  nonlinear
sigma models with double-derivative interactions.
Such theories are super-renormalizable: though certain one- and two-loop 
Feynman diagrams are superficially divergent and regularization is necessary. 
However, there is no need to renormalize infinities away because 
the infinities of different graphs cancel each other 
and quantum mechanics is finite.
Different regularization prescriptions give in general
different finite answers for the same Feynman diagram. 
This situation is rather familiar in QFT: it simply means that 
there are free parameters entering in the theory (which are 
equivalent to the ordering ambiguities of canonical quantization)
that can only be fixed by requiring further constraints 
(finite renormalization conditions).
The latter simply parametrize different physical phenomena
which can be described by the quantum mechanical model under consideration.

It is sometimes claimed that one does not need any 
``artificial'' counterterm at all because the theory 
has no divergences.
As the results of this article show, in all regularization
schemes studied so far one always needs finite local counterterms.
In fact, finite local counterterms are in general to be expected 
because they just amount to finite additive renormalizations 
needed to implement the renormalization conditions.

In the recent past, two different regularization schemes 
for nonlinear sigma models on a finite time interval
have been discussed carefully: mode regularization 
\cite{Bastianelli:1992be,Bastianelli:1993ct,Bastianelli:1998jm}
and time discretization \cite{Bastianelli:1998jm, deBoer:1995hv, Schalm:1998}.
A detailed comparison carried out to three loops 
shows that both schemes produce the same physics \cite{Bastianelli:1999jb}.
However, they both break manifest general coordinate invariance
at intermediate stages and require noncovariant counterterms 
to restore that symmetry in the final result.
It is important to stress  that these counterterms are 
unambiguously determined in each scheme.
Nevertheless, lack of manifest covariance is annoying and constitutes 
a technical limitation: at higher loops one must expand the 
non-covariant counterterms to get the corresponding vertices but one cannot
employ covariant techniques to simplify that computation.

Recently, dimensional regularization has been employed to define 
a new regulated version of the path integral for an 
infinite time interval~\cite{Kleinert:1999aq}. 
By evaluating the partition function of a particular massive 
nonlinear sigma model with
a one-dimensional  flat target space, it was found that no noncovariant 
counterterms were needed to obtain the correct result.  Since target space 
in~\cite{Kleinert:1999aq} 
was only one-dimensional, covariant counterterms could not be detected since 
these are proportional to the scalar curvature $R$. 
It is the purpose of this letter to extend the proposal of 
ref.~\cite{Kleinert:1999aq} to a higher dimensional target space
and to demonstrate that a covariant counterterm
is needed.
This counterterm turns out to be $V_{DR} = {\hbar^2\over 8}R$.

Let us present first a discussion on the limits of dimensional 
regularization applied to quantum mechanics as used in \cite{Kleinert:1999aq}.
The main problem is that it seems to require an infinite propagation time. 
In fact, one obtains a continuum momentum space (the energy in one dimension)
only upon Fourier transforming the infinite time dimension.
Integrals in momentum space are
regulated dimensionally afterwards \cite{'tHooft:1972fi}.
Instead, it would be desirable to regulate and 
compute the path integral for a finite propagation time.
The latter could be interpreted as a proper time, 
thus making it useful for relativistic applications 
in the world line approach to QFT~\cite{wla}. 
A related problem is that the infinite propagation time 
introduces infrared divergences in massless models, and
requires a harmonic term as infrared regulator. In ref.~\cite{Kleinert:1999aq} 
only a massive model was considered. 
The harmonic term ruins general coordinate invariance:
a potential of the form $V \sim \omega^2 g_{ij}(x)x^i x^j$
is not a scalar since the coordinates $x^i$ do not transform
as the components of a vector. 
Invariance in the final result could be recovered
in the limit $\omega \rightarrow 0$ if the propagation time
would be kept finite, but that limit is not possible
in the dimensional regularization described above which requires
an infinite propagation time.
%Invariance in the final result cannot be recovered 
%even in the limit $\omega \rightarrow 0$.
Given that general coordinate invariance is necessarily softly broken,
one may use as well a potential
$V \sim \omega^2 g_{ij}(0)x^i x^j$ 
as infrared regulator. The latter is quadratic even far away 
from the origin of the chosen coordinate system and will not modify the 
interaction vertices. This soft breaking of general coordinate 
invariance is not expected to 
modify the counterterm $V_{CT}$ since such a counterterm is sensitive only 
to the ambiguities due to ultraviolet divergences.

We now proceed to
test the proposal of ref.~\cite{Kleinert:1999aq} 
in a class of sufficiently general models 
and relate it to the other regularization methods mentioned above.
The calculation in \cite{Kleinert:1999aq} is enough to indicate
that possible counterterms will be covariant, but since it involves
a single coordinate it misses terms proportional to the curvature.
Our strategy will be to compute terms in the effective action using both 
mode regularization (MR) and dimensional
regularization (DR). Equating the results fixes the counterterm needed in 
dimensional regularization to be $V_{DR} = {\hbar^2\over 8}R$.

First, let us briefly review some known 
facts.
Quantization of a free particle on a curved space 
produces in the quantum Hamiltonian $\hat H$
an undetermined term proportional to the scalar curvature,
$\hat H = - {\hbar^2\over 2} \Delta + \alpha\hbar^2 R $.
This is easily seen using canonical (operatorial) methods:
ordering ambiguities are encountered in the construction of 
the quantum Hamiltonian from the classical one and give rise 
to terms with at most two derivatives on the metric. Then, 
requiring general coordinate invariance leaves only a term proportional
to the scalar curvature.
Using path integrals this arbitrary coupling will appear
as a correction to the effective action proportional
to the scalar curvature
\bea 
\la 0|e^{-\beta\hat H/\hbar}|0\ra =
\int {\cal D}x\ {\rm e}^{-S[x]/\hbar} = {\rm e}^{- \Gamma/\hbar} =
{\rm exp} \left \{ - {1\over \hbar}\int^{\beta}_{0} 
dt \left [ \cdots +
(\alpha + {1\over 12})\hbar^2 R +\cdots] \right ] \right \}
\label{effe}
\ena
where the first equality reminds us of the equivalence of
canonical and path integral quantization 
($|0\ra$ and $\la 0|$ are eigenstates of the position 
operator $\hat x$ with eigenvalue zero)
and in the second equality we have the definition of the effective 
action $\Gamma$. The term ${\hbar^2\over 12}R$ is partially due to the
counterterm and partially due to two-loop diagrams, see eq.~(\ref{amp}). 
Henceforth we set $\hbar =1$.

We are going to compute the corrections to the effective  
action $\Gamma$ as function of the various couplings using both 
mode and dimensional regularization.
In the former we can be general and allow for a finite
propagation time $\beta$. Then we take the limit 
$\beta \rightarrow \infty$, which is safe in the presence of an infrared 
regulator, and compare the result with dimensional regularization.
It is known that the former requires the counterterm 
\bea
V_{MR}={1\over 8} R  -{1\over 24} g^{ij} g^{kl} g_{mn}
\Gamma_{ik}^m \Gamma_{jl}^n 
\label{vmr}
\eea 
to produce a general coordinate invariant result with $\alpha =0$
\cite{Bastianelli:1998jm}.
We will see that dimensional regularization
will match the result when using a counterterm 
\bea
V_{DR} = {1\over 8}R
\label{vdr}
\eea
which is manifestly covariant. For comparison we mention that the counterterm 
for time-slicing, needed to obtain the same result as mode regularization,
is different, see eq.~(\ref{vts}).
%
%
%
%%%%% The 3-loop calculations with Riemann normal coordinates
%
%
\section{The 3-loop calculation with Riemann normal coordinates}
The model we analyze is given by
\bea
S[x^i] = \int_{t_i}^{t_f} \!\!\! 
dt\ \biggl [ {1\over 2} g_{ij}(x)\dot x^i \dot x^j 
+ {1\over 2} \omega^2 g_{ij}(0) x^i  x^j +  V_{CT}
\biggr ]
\label{model}
\ena
where $\omega$ is a frequency needed as infrared regulator
and $V_{CT}$ is the counterterm for the regularization scheme chosen.
Using Riemann normal coordinates, we will 
need to compute up to three loops since the noncovariant part of the 
counterterm~(\ref{vmr}), when expanded around the origin of the coordinates, 
only gives contributions from 3~loops onwards.
We want to make sure that 
noncovariant counterterms are not required when using dimensional
regularization, as noticed in ref. \cite{Kleinert:1999aq}.
In the next section we shall repeat the calculation below for general
coordinates but with only two-loop graphs. Since in general coordinates the
first derivatives of the metric do not vanish, we get nonvanishing 
contributions from the noncovariant parts of~(\ref{vmr}) already 
at the two-loop level. This gives an additional 
nontrivial check on the covariance of the counterterm of dimensional
regularization on the infinite time interval.

The counterterm is effectively of order $\hbar^2$ since
it first appears at two loops, but for notational 
convenience we are using units where $\hbar=1$.
As already mentioned, the harmonic potential breaks general 
coordinate invariance since it selects those coordinates
in which the potential is quadratic.
We have chosen them to be Riemann normal coordinates
as a definition of our model, so that
the metric has the expansion
\bea
 g_{ij}(x) &=&
\delta_{ij} + {1\over 3} R_{kijl}(0) x^k x^l
+{1\over 6} \nabla_m R_{kijl}(0)
x^k x^l x^m 
\nonumber
\\
&& + \biggl(
{1\over{20}} \nabla_m \nabla_n R_{kijl}(0) +
{2\over{45}} R_{kipl} R_{mj}{}^p{}_n (0)
\biggr ) x^k x^l x^m x^n + O(x^5).
\ena

We find it convenient to use a rescaled time parameter
$\tau$ with $t = \beta \tau + t_f$
and $\beta= t_f- t_i$, so that $-1 \leq\tau \leq 0$.
An infinite propagation time will be recovered in the limit 
$\beta \rightarrow \infty$,
while  for finite $\beta$ this setting allows us
to compare easily with the results for $\omega=0$
which were reported in \cite{Bastianelli:1999jb} 
using similar notations \footnote{
Our conventions follow from 
$ [\nabla_i,\nabla_j ] V^k = R_{ij}{}^k{}_l V^l$, $R_{ij}=R_{ik}{}^k{}_j$.
Thus, the scalar curvature $R=R_i{}^i$ of a sphere is negative.}.

With this rescaling, and introducing the ghost $a^i,b^i,c^i$ 
for a correct treatment of the measure
\cite{Bastianelli:1992be,Bastianelli:1993ct}, 
we aim to compute the following path integral
with two different regularization schemes, 
mode regularization (MR) and dimensional regularization (DR),
\bea
\int 
 {\cal D} x  {\cal D} a  {\cal D} b  {\cal D} c
\ {\rm e}^{- {1\over\beta} S }
\label{pi}
\eea
with
\bea
&&\hskip -1cm
S \equiv S[x,a,b,c] =
\int_{-1}^{0} \! \! \! d\tau \ \biggl (
{1\over 2} g_{ij}(x) (\dot x^i \dot x^j + a^i a^j +
b^i c^j) +{1\over 2} (\beta\omega)^2 g_{ij}(0) x^i x^j +
\beta^2 V_{CT}(x) \biggr )
\label{seven}
\eea                 
and with the boundary conditions that all fields vanish at $t=t_i, t_f$,
(i.e. at $\tau =-1,0$).

For the perturbative evaluation (in the coupling constants contained in
the metric $g_{ij}(x)$) 
it is convenient to split the action into a quadratic part
$S_2$ and an interacting part $S_{int} = S_3+S_4+S_5+S_6+\cdots$
\bea
S_2 &=&\int_{-1}^{0}   \!\!\! d\tau \ 
\biggl [{1\over 2} \delta_{ij}  
(\dot x^i \dot x^j +a^i a^j +b^i c^j) + {1\over 2} \delta_{ij}
(\beta\omega)^2 x^i x^j \biggr ] \\
S_3 &=& 0 \\
S_4 &=& \int_{-1}^{0}   \!\!\! d\tau \ \biggl [
{1\over 6} R_{kijl} x^k x^l
(\dot x^i \dot x^j + a^i a^j +b^i c^j) + 
\beta^2 V_{CT} \biggr ] 
\\
S_5 &=& \int_{-1}^{0}  \!\!\!  d\tau \ \biggl [
{1\over 12} \nabla_m R_{kijl}
x^k x^l x^m (\dot x^i \dot x^j +a^i a^j +b^i c^j) 
+ \beta^2  x^i \partial_i  V_{CT} \biggr ] 
\\
S_6 &=& \int_{-1}^{0}   \!\!\!  d\tau \  \biggl [
\biggl(
{1\over{40}} \nabla_m \nabla_n R_{kijl} +
 {1\over{45}} R_{kipl} R_{mj}{}^p{}_n
\biggr )
x^k x^l x^m x^n 
(\dot x^i \dot x^j +a^i a^j +b^i c^j) 
\nonumber
\\ 
&& \hskip 1.3cm 
+{\beta^2\over 2} x^i x^j \partial_i \partial_j  V_{CT} \biggr ] .
\eea
Note that all structures like $R_{ijkl}$, $V_{CT}$
and derivatives thereof are evaluated at the origin of the Riemann
coordinate system, but for notational simplicity we do not indicate
so explicitly from now on.

From $S_2$ one recognizes the propagators
\bea
\la x^i(\tau) x^j(\sigma)\ra &=&
-\beta\ \delta^{ij}\ \del(\tau,\sigma)
\nonumber\\
\la a^i(\tau) a^j(\sigma)\ra &=&  \beta\ \delta^{ij}\
\Delta_{gh}(\tau, \sigma) \label{xx}
\\
\la b^i(\tau) c^j(\sigma)\ra &=& -2\beta\ \delta^{ij}\
\Delta_{gh}(\tau,\sigma)
\nonumber
\eea
where the functions $\del(\tau,\sigma)$,
$\Delta_{gh}(\tau,\sigma)$ are to be defined shortly in each 
regularization scheme.
Then, the transition element, eq. (\ref{effe}),
at three loops is given by
\bea
{\cal Z}={\rm e}^{- \Gamma} =
 A\: {\rm exp}\biggl \{ \biggl \langle -
{1\over \beta }(S_4 + S_5 + S_6) \biggr \ra +
\biggl \langle {1\over  2\beta^2}  S_4^2  
\biggr \ra_{con} + \cdots \biggr \} 
\label{perturb}
\eea
where the subscript $\it con$ refers to connected diagrams only.
The constant $A$ is the normalization of the exact path integral
for $S_2$ which describes a harmonic oscillator
in $D$ dimensions~\cite{Feyn,somerefs} 
\bea
A= \biggl ( {\omega\over {2 \pi \sinh(\beta\omega)}}\biggr)^{D\over 2}.
\eea
For $\omega=0$ this term becomes the familiar Feynman
measure for a free particle $(2\pi \beta)^{-D/2}$.
The perturbative contributions are obtained by computing the various 
Wick contractions. We record the results in terms of
$\del$ and $\del_{gh}$ through three loops;  
the symbol * denotes counterterms. 
The nonzero contributions are
\bea
&&\hspace{-0.75cm}\biggl\langle - {1\over \beta} S_4 \biggr \rangle =
\raisebox{-1.75ex}{\includegraphics*[0cm,0cm][3cm,1.5cm]{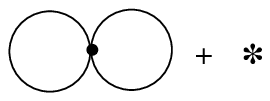}}
=- I_2\ {\beta\over 6} R  -\beta V_{CT} \\
%\biggl \langle - {1\over \beta} S_5 \biggr \rangle &=&
%\raisebox{-3.75ex}{\includegraphics*[0cm,2cm][3.75cm,4cm]{f4.eps}} 
%=0 \\
&&\hspace{-0.75cm}\biggl \langle - {1\over \beta} S_6 \biggr \rangle = 
 \raisebox{-2.5ex}{\includegraphics*[0cm,0cm][2.75cm,1.5cm]{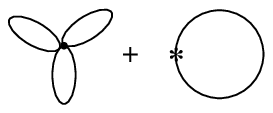}}
= I_8\ \beta^2 \biggl ({1\over 20}
\nabla^2 R + {1\over 45} R_{ij}^2
+  {1\over 30} R_{ijmn}^2 \biggr )
+ I_9\ {\beta^2\over 2}\partial^i \partial_i V_{CT}
\label{clover}
\\
&&\hspace{-0.75cm}\biggl \langle {1\over 2 \beta^2} S_4^2 \biggr
\rangle_{\! con}
= \raisebox{-2ex}{\includegraphics*[0cm,0cm][3.95cm,1.5cm]{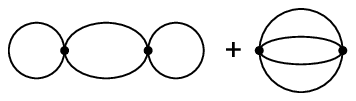}}= 
I_{14}\ {\beta^2 \over 36} R_{ij}^2 + I_{15}\ {\beta^2 \over 24} R_{ijmn}^2 
\eea
being $\la S_5 \ra$ proportional to at least one classical field that is zero
since $x_i=x_f=0$. The integrals $I_n$ are  given by
\bea    
I_2 &=&
\int_{-1}^{0} \!\!\! d\tau \ \bigl (
\del \ (\ddeld + \Delta_{gh}) - \ddel^2 \bigr )|_\tau 
\label{20}
\\
I_{8} &=&
\int_{-1}^{0}  \!\!\! d\tau \
\bigl ( \del^2\ (\ddeld + \Delta_{gh}) - \ddel^2\ \del \bigr )|_\tau 
\\
I_{9} &=&
\int_{-1}^{0}  \!\!\!  d\tau \
\del|_\tau 
\\            
I_{14} &=&
%\idtds 
\int_{-1}^{0}  \!\!\!  d\tau \! \int_{-1}^{0}  \!\!\!  d\sigma
\biggl (
\del|_\tau \ (\ddeld{}^2 - \Delta_{gh}^2 )\  \del|_\sigma
- 4\ \del|_\tau \ \ddeld\ \ddel\ \deld|_\sigma
\nonumber \\
&&+
2\ \del|_\tau \ \ddel^2\  (\ddeld + \Delta_{gh})|_\sigma
+
2\ \deld|_\tau\ \del\ \ddeld\  \deld|_\sigma
+
2\ \deld|_\tau\ \ddel\ \deld\ \deld|_\sigma
\nonumber \\
&&-
4\ \deld|_\tau \ \del\ \ddel\ (\ddeld + \Delta_{gh})|_\sigma
+
(\ddeld + \Delta_{gh})|_\tau \ \del^2 \ (\ddeld + \Delta_{gh})|_\sigma
\biggr ) 
%\nonumber
\\               
I_{15}
&=& \int_{-1}^{0}  \!\!\!  d\tau \! \int_{-1}^{0}  \!\!\!  d\sigma
% \idtds 
\biggl (
\del^2 \ (\ddeld{}^2 - \Delta_{gh}^2)
+
\ddel^2 \ \deld{}^2
%\nonumber \\
%&-&
-2\ \del\ \ddel\ \deld\ \ddeld
\biggr )  .
\label{24}
\eea               
We have kept the same names and notations for the integrals $I_n$ as in
\cite{Bastianelli:1999jb}
to facilitate comparison for the limit $\omega \rightarrow 0$
possible in mode regularization when $\beta$ is kept finite.
We recall that $\del|_{\tau}\equiv\del(\tau,\tau)$
and $\ddel\equiv{\partial\over \partial\tau}\del(\tau,\sigma)$ while
$\deld\equiv{\partial\over \partial \sigma}\del(\tau,\sigma)$.

Let us first consider mode regularization.
Here one expands all fields in a Fourier sine series and keeps 
all modes up to a large mode number $M$. 
The limit $M\rightarrow \infty$ is taken 
after having computed all integrals.
In practice, one manipulates the integrals by partial integration
to put them into a form which can be computed directly and without 
ambiguities in the continuum. 
One partially integrates such that all double derivatives of 
$\del$, namely $\ddeld$ and  $\Delta_{gh}\equiv \dddel_0$ 
are removed. 
If this is not possible, one casts the expressions in a form such 
that the integrands vanish at the end-points. 
In the latter case, singularities like $\delta(\tau)$ and $\delta(\tau +1)$ 
are neutralized.
With this prescription one recognizes that
the function $ \del(\tau,\sigma)$ appearing in the propagator
is given  by
\bea
\del(\tau,\sigma) = \sum_{m=1}^{M} \biggl [
{-2\over {(\pi m)^2 +(\beta \omega)^2}} {\rm \sin}(\pi m\tau)
{\rm \sin}(\pi m\sigma)\biggr ]
\eea     
while as anticipated 
one can represent $ \Delta_{gh}(\tau,\sigma)=\dddel_0(\tau,\sigma) $
with
\bea
\del_0(\tau,\sigma) 
= \sum_{m=1}^{M} \biggl [
{-2\over {(\pi m)^2}} {\rm \sin}(\pi m\tau)
{\rm \sin}(\pi m\sigma)\biggr ] .
\eea                         
Their continuum limit ($M\rightarrow \infty$) is given by
\bea
\del(\tau,\sigma)&=&{1\over{ \beta\omega \sinh(\beta\omega)}}
\biggr [\theta(\tau -\sigma)
\sinh(\beta\omega \tau) \sinh \bigl(\beta\omega (\sigma+1)\bigr)+
\nonumber \\
&&\hspace{2.5cm}+\theta(\sigma -\tau)
\sinh(\beta\omega \sigma) \sinh \bigl(\beta\omega (\tau+1)\bigr)
\biggl]
\label{27}
\\
\Delta_{gh}(\tau,\sigma) 
&=& \delta(\tau -\sigma)  . 
\eea
It is easy to check that $\left[\partial_\tau^2-(\beta\omega)^2\right]
\del(\tau,\sigma)=\delta(\tau-\sigma)$ and $\del(0,\sigma)=\del(-1,\sigma)
=\del(\tau,0)=\del(\tau,-1)=0$.

Now, we can compute the various $I_n$ and obtain the results summarized 
in Table~\ref{tab}, 
\begin{table}
\begin{center}
\[\begin{array}{|c|c|c|c|c|} \hline 
I_2 & I_8 & I_9 & I_{14} & I_{15}\\ \hline
\raisebox{-0.1ex}{$-{1\over 4}$} 
& -\raisebox{-0.1ex}{${1\over 8}I(\beta\omega)$} &
\raisebox{-0.1ex}{$ {1\over 2}I(\beta\omega)$} 
& \raisebox{-0.1ex}{${1\over 4}I(\beta\omega)$}
&\raisebox{-0.1ex}{$ {1\over 6}I(\beta\omega)$}\\[0.1cm] \hline
\end{array}\]
\end{center}
\vspace{-0.5cm}
\caption{Results in mode regularization at finite $\beta$}\label{tab}
%\vspace{-0.5cm}
\end{table}
where we have found it convenient to define the function
\bea 
I(a)= {1-a \coth(a) \over a^2} .
\eea
As an example how these results are obtained, consider the ``clover leaf'' 
graph in (\ref{clover}) corresponding to $I_8$. Using that in mode 
regularization 
\bea
(\ddeld + \Delta_{gh})|_\tau = \partial_\tau ((\ddel) |_\tau)-
(\beta\omega)^2 \del |_\tau,
\eea 
the first term in $I_8$ yields 
\bea
\int_{-1}^{0}  \!\!\!  d\tau \
( -4\ddel^2 \del-(\beta\omega)^2 \del^3 )|_\tau .
\label{I81}
\eea
Hence 
\bea
I_8 = \int_{-1}^{0}  \!\!\!  d\tau \
(-5 \ddel^2 \del -(\beta\omega)^2 
\del^3 )|_\tau .
\label{I82}
\eea
Then from eq. (\ref{27}) we obtain
\bea
\del|_\tau &=& { {{\rm sinh}(\beta\omega\tau) {\rm sinh}(\beta\omega(\tau+1))}
\over 
{ \beta\omega {\rm sinh}(\beta\omega)} } \\
(\ddel) |_\tau &=& { {{\rm sinh}(\beta\omega(2\tau +1))}\over 
{2 {\rm sinh}(\beta\omega)}} 
\eea
and substitution into (\ref{I82}) yields the result for $I_8$ 
as given in Table~\ref{tab}.

In this regularization scheme the counterterm to be used is $V_{MR}$
as given in eq.~(\ref{vmr}).  When evaluated at the origin
of the Riemann normal coordinates it produces 
\bea
&& V_{MR} = {1\over 8} R 
\\
&& \partial^i \partial_i V_{MR} = {1\over 8} 
\nabla^2 R -{1\over 36} R_{ijkl}R^{ijkl}  .
\eea

As an aside, we can check the correctness of the $\omega\rightarrow 0$
limit. Since
$ I(\beta\omega)\rightarrow-{1\over 3}$ for $\omega\rightarrow 0$,
one can verify that the results in \cite{Bastianelli:1999jb}
are reproduced
\bea 
{\cal Z} = A\: {\rm exp}\biggl\{- \biggl [ \beta {1\over 12} R + 
\beta^2 \biggl ( {1\over 120} \nabla^2 R
+{1\over 720}  R_{ij}^2  -{1\over 720}  R_{ijkl}^2 
\biggr ) + \cdots \biggr]\biggr\}.
\label{amp}
\eea
This result is expected to be covariant 
\cite{Bastianelli:1998jm,Bastianelli:1999jb} and the use
of Riemann normal coordinates shows immediately which is the
covariant form of the effective action.

On the other hand, for $\omega\neq 0$ and $\beta \rightarrow \infty$ 
one gets $\beta I(\beta\omega)\rightarrow-{1\over \omega}$, 
and thus
\bea
{\cal Z} =  
A\: {\rm exp}\biggl\{- \beta\biggl [ {1\over 12} R + {1\over \omega}
\biggl ( {1\over 40} \nabla^2 R + 
{1\over 240} R_{ij}^2 - {1\over 240} R_{ijkl}^2 \biggr) 
+\cdots \biggr]\biggr\}  .
\label{free}
\eea
Now, this result in not expected to be covariant because of the presence 
of the mass term $\omega$. The apparent covariance of (\ref{free})
is just a coordinate artifact of the Riemann normal coordinates
(this point will be self-evident in the calculations of the next section).
The result (\ref{free})  
is what one should obtain in dimensional regularization as well.

Thus, let us turn to dimensional regularization.
The propagators are represented as in (\ref{xx}) with
\bea
\Delta(\tau,\sigma) &=& -{1\over \beta}
\int {dk\over 2\pi} {{\rm e}^{-i k \beta(\tau-\sigma)} 
\over{ k^2 +\omega^2}} \\
\Delta_{gh}(\tau,\sigma) &=& -{1\over \beta}
\int {dk\over 2\pi} {\rm e}^{-i k \beta(\tau-\sigma)} \ .
\eea
Note that, strictly speaking, one should use 
an infinite $\beta$, which anyway cancels in (\ref{xx}), and a
finite $t \equiv \beta \tau$ and $s \equiv \beta\sigma$. 
Now one can use dimensional regularization to compute the 
various integrals (with  momenta contracted as suggested
by the kinetic term continued to $D$ dimensions) and then
take the limit $D \rightarrow 1$.  
Using the formulas given in \cite{Kleinert:1999aq}
(and also in \cite{Kleinert:1999ei} where dimensional regularization
is used in configuration space),
one recognizes that the ghosts are effectively regulated to give 
a vanishing contribution
(this is due to the fact that $\delta^{(n)}(0)$
is zero in dimensional regularization),
while the remaining integrals 
give the results summarized in Table~\ref{tab2}.

\begin{table}
\begin{center}
\[\begin{array}{|c|c|c|c|c|} \hline 
I_2 & I_8 & I_9 & I_{14} & I_{15}\\ \hline
\raisebox{-0.1ex}{$-{1\over 4}$} 
&\raisebox{-0.1ex}{$ {1\over 8\beta\omega}$} 
& \raisebox{-0.1ex}{$-{1\over 2\beta\omega}$} 
&\raisebox{-0.1ex}{$ -{1\over 4\beta\omega}$}
&\raisebox{-0.1ex}{$ 0$}\\[0.1cm] \hline
\end{array}\]
\end{center}
\vspace{-0.5cm}
\caption{Results in dimensional regularization at $\beta=\infty$}\label{tab2}
%\vspace{-0.5cm}
\end{table}
 
It is immediate to verify that the result (\ref{free})
is reproduced once one uses the counterterm $V_{DR}={1\over 8} R$
(of course, in the limit of infinite $\beta$ this result
is unaffected by the infrared divergence related to the infinite time 
integral and remains finite). Thus, 
we conclude that $V_{DR}={1\over 8} R$
is the counterterm needed in dimensional regularization
to have $\alpha =0$ in eq. (\ref{effe}).

Of course, we could have compared as well 
dimensional regularization with time
slicing regularization \cite{deBoer:1995hv}
and obtain the same result.
In that case, one should remember that
time slicing (TS) requires different rules to compute
the integrals in eqs.~(\ref{20}-\ref{24})
but also a different counterterm \cite{deBoer:1995hv,Gervais:1976ws}
\bea
V_{TS}={1\over 8} R  +{1\over 8} g^{ij} \Gamma_{ik}^l \Gamma_{jl}^k. 
\label{vts}
\ena
As an extra check, in what follows we also verify the necessity 
of the counterterm
$V_{DR}$ at two loops but using arbitrarily chosen coordinates.
%
%
%
%the two-loop calculation with general coordinates===========================
%
%
\section{The two-loop calculation with general coordinates}
In this section we repeat the calculation for the amplitude (\ref{effe})
 using general
coordinates going as far as two loops. Again we perform the calculation using 
mode regularization along with the counterterm (\ref{vmr}) and dimensional 
regularization with the counterterm (\ref{vdr}) 
applied to the model (\ref{model}) where $x^i$ are {\it now} 
general coordinates. 
Writing (\ref{vmr}) explicitly in terms of the metric tensor
\bea
V_{MR}&=&{1\over 8} R  -{1\over 24} g^{ij} g^{kl} g_{mn}
\Gamma_{ik}^m \Gamma_{jl}^n = {1\over 8} R-{1\over 32}
\left (
\partial_i g_{jk}\right )^2 +{1\over 48} \left (
\partial_i g_{jk}\right )\left (
\partial_j g_{ik}\right )
\label{vmr1}
\eea 
makes it clear that one will get nonzero contribution from the
noncovariant parts of the counterterms already at the two-loop level. 
Indeed the 
derivatives of the metric do not vanish at the origin of an arbitrary system
of coordinates contrarily to what happens in Riemann normal coordinates where
 they do vanish.   
The expansion of the metric $g_{ij}(x)$ around the origin gives the same 
quadratic action of the 
previous section and thus the propagators are the same as well. 
The interacting part is $S_{int}
=S_3+S_4+\cdots$, being
\bea
S_3 &=&\int^0_{-1}d\tau\: {1\over 2}\partial_k g_{ij}x^k 
\left(\dot x^i \dot x^j + a^i a^j 
+ b^i c^j\right)\\
S_4 &=&\int^0_{-1}d\tau \left[{1\over 4}\partial_k
 \partial_l g_{ij}x^k x^l \left(\dot x^i \dot x^j + a^i a^j 
+ b^i c^j\right) + \beta^2 V_{CT}\right]
\eea
where metric and derivative thereof and $V_{CT}$ are evaluated 
at the origin of the system of coordinates. The transition element at 
two-loop is given by
\bea
{\cal Z}=A\: {\rm exp}\left\{ \left\la 
-{1\over \beta}\left(S_3+S_4\right)\right\ra 
+\left\la {1\over 2\beta^2}S_3^2 \right\ra_{con} +\cdots \right\}
\eea 
where $\la S_3 \ra$ vanishes because it contains an odd number 
of quantum fields while 
\bea
&& 
\hskip -1.3cm
\biggl \langle - {1\over \beta} S_4 \biggr \rangle =
-{\beta\over 4} \biggl [  A_1 \partial^2 g  
+ 2 A_2  \partial^j g_j  \biggr ]
 -\beta V_{CT} 
\\
&& \hskip -1.3cm
\biggl \langle  {1\over 2\beta^2} S_4 \biggr \rangle_{con} \!\! = 
- {\beta\over 8} \biggl [ B_1 (\partial_i g)^2 
+4 B_2  (\partial_j g) g^j 
+ 2 B_3 (\partial_i g_{jk})^2 
+4 B_4 (\partial_i g_{jk})\partial_j g_{ik}
+ 4 B_5 g_j^2
\biggr ].
\eea
We have used the shorthand notation: 
$\partial^2 g\equiv g^{ij}g^{kl}\partial_k\partial_l g_{ij}$, 
$\partial_k g\equiv g^{ij}\partial_k g_{ij}$, 
$g_k\equiv g^{ij}\partial_i g_{jk}$,
$\partial^j g_j\equiv g^{ik}g^{jl}\partial_k\partial_l g_{ij}$.
The results obtained  from this calculation are summarized in
Table~\ref{tab:2lo}, where the column 
``Result'' refers to the computations done using dimensional 
regularization (DR) and mode regularization (MR) of the integrals 
shown in the column aside. In the same line 
we also report a pictorial representation and the ``tensor structure''
associated to each diagram. 
Recalling that the scalar curvature is given by
\bea
R=\partial^2 g -\partial^j g_j -{3\over 4}\left(\partial_k g_{ij}\right)^2
+{1\over 2}\left(\partial_i g_{jk}\right)\partial_j g_{ik}
+{1\over 4}\left(\partial_j g\right)^2 
-\left(\partial_j g\right)g^j +g_j^2
\eea        
and using the results from Table~\ref{tab:2lo}, 
the amplitude ${\cal Z}$ reads
\bea
{\cal Z}=
A\:{\rm exp}\left\{-{\beta \over 16}\partial^2 g
+{\beta \over 8}\partial^j g_j +{\beta\over 48}(\partial_i g_{jk})^2
-{\beta\over 12}(\partial_i g_{jk})\partial_j g_{ik}
+{\beta \over 8}(\partial_j g)g^j-{\beta \over 8}g_j^2
\right\}
%\nonumber\\
%&=&A\:{\rm exp}\left\{-{\beta \over 8}R+{\beta \over 16}\partial^2 g
%-{\beta\over 48}(\partial_i g_{jk})\partial_j g_{ik}
%-{7\beta\over 96}(\partial_i g_{jk})^2+{\beta\over 32}\left(\partial_j 
%g\right)\partial^j g\right\}
\label{amp:2lo}
\eea
for both regularization schemes.
Therefore, also in this case, dimensional regularization yields the same
transition amplitude as mode regularization
only requiring the covariant counterterm ${1\over 8}R$.

%\vspace{-2cm} 
\begin{table}
\begin{center}
\vspace{-2cm}
\begin{tabular}{|l|l|c|l|} \hline 
\emph{Integral} & \emph{Result} & \emph{Diagram} & \emph{Tensor} \\ 
%\cline{1-1}\cline{3-3}
& \emph{{\rm DR[MR]}} & & \emph{structure}\\ \hline
$A_1\equiv\int_{-1}^0\del |_\tau(\ddeld+\del_{gh}) |_\tau$  & 
$-{1\over 4} \:\:\:\left [-{1\over 4} 
\right ]$ 
&$\raisebox{-1ex}{\includegraphics*[-0.5cm,0cm][2.5cm,2.5cm]{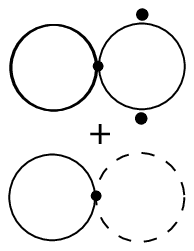}}$ & 
$\partial^2 g$  \\ \cline{1-4}
$A_2\equiv\int_{-1}^0 \left (\ddel |_\tau \right )^2 $ & $0 \:\:\:\left [0 
\right ]$ 
&$\raisebox{-1ex}{\includegraphics*[-0.5cm,0cm][2.5cm,1.2cm]{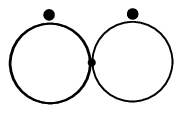}}$ 
&$\partial^j g_j$ \\ \hline
$B_3\equiv\int_{-1}^0\int_{-1}^0 
\del(\ddeld{}^2 -\del_{gh}^2)$ & $ {7\over 24} 
\:\:\: \left [{5\over 12}
\right ] $ 
&$\raisebox{-1ex}{\includegraphics*[-0.5cm,0cm][2.5cm,1cm]{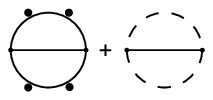}}$ &
$ (\partial_i g_{jk})^2 $ \\ \hline
$B_4\equiv\int_{-1}^0\int_{-1}^0 \left(\ddeld\right) 
\deld \left(\ddel\right) $ & 
$ {1\over 24} \:\:\:
\left [0
\right ]$ &$\raisebox{-1ex}{\includegraphics*[-1cm,0cm][2cm,1cm]{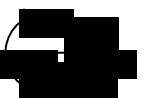}}$ &
$(\partial_i g_{jk})\partial_j g_{ik} $ \\ \hline
$B_5\equiv\int_{-1}^0\int_{-1}^0 
\ddel |_\tau\left(\ddeld\right)\deld |_\sigma$ & 
$ 0 \:\:\:
\left [0
\right ]$ &$\raisebox{-1ex}{\includegraphics*[-0.5cm,0cm][3cm,1cm]{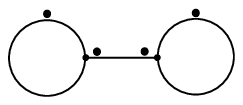}}$ & 
$g_j^2$ \\ \hline
$B_2\equiv\int_{-1}^0\int_{-1}^0 (\ddeld
+\del_{gh}) |_\tau\deld\left(\deld |_\sigma\right)$ & $0 
\:\:\:\left 
[0 \right ]$ 
&$\raisebox{-1ex}{\includegraphics*[0cm,0cm][3cm,2.35cm]{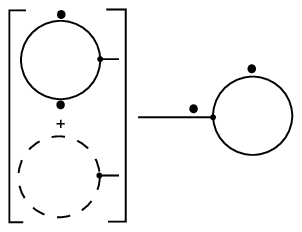}}$ & 
$(\partial_j g)g^j$ \\ \hline
$B_1\equiv\int_{-1}^0\int_{-1}^0 (\ddeld+\del_{gh}) |_\tau\del(\ddeld
+\del_{gh}) |_\sigma$ & $-{1\over 4} 
\:\:\:\left [-{1\over 4}\right ]$ &
$\raisebox{-1ex}{\includegraphics*[-0.25cm,0cm][3.5cm,2.25cm]{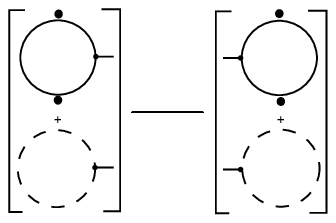}}$ 
& $(\partial_j g)^2 $ 
\\ \hline   
\end{tabular}
\end{center}
\vspace{-0.5cm}
\caption{2-loop results with dimensional and mode 
regularization}\label{tab:2lo}
\end{table}

Note that in Riemann normal coordinates
$\partial^2 g={2\over 3}R$ and 
$\partial^j g_j=-{1\over 3}R$ at the origin; substituting these identities,
(\ref{amp:2lo}) reduces to the two-loop
part~of~(\ref{amp}).  Obviously the result is not covariant
as the covariance of the model in eq. (\ref{model}) is explicitly
broken by the mass term and cannot be recovered even in the limit
$\omega\rightarrow 0$ since~(\ref{amp:2lo}) is $\omega$ independent. 
Therefore dimensional regularization of the path integral 
on an infinite time interval does not preserve target space  
general covariance, contrarily to what stated in~\cite{Kleinert:1999aq}. 

%\vspace{-2cm}
%
%
%
%%%%%%%%%%%%%%%Conclusions
%
%
%\vspace{5cm}
\section{Conclusions}
In this letter we considered quantum mechanical path integrals in curved 
space with an infinite propagation time. We computed transition amplitudes 
both using dimensional regularization (DR) and other (in
this context more established) regularization schemes. We showed that DR 
does not need noncovariant counterterms in order to 
reproduce the correct answer, as already noticed in a simpler model
 in~\cite{Kleinert:1999aq}, but it 
does need a covariant two-loop counterterm, namely $V_{DR}={\hbar^2\over 8}R$. 
We took an infinite propagation  time in order to have a 
continuous momentum spectrum and to be able to use DR in the  usual way. 
This forced us to add an infrared
regulator: a mass term. The unpleasant feature of this term is that it 
 breaks manifest general covariance. 
Furthermore, for applications to quantum field theories such as computations 
of anomalies, one needs path integrals
on a finite time interval. We are at present working on an approach to use
 dimensional regularization at finite 
$\beta$. The crucial question is whether again only covariant counterterms
 are needed.

%\newpage
%%%%%%%%%%%%%%%%%%%%%%%%%%%%%%%%%%%%%%%%%%%%%%%%%%%%%%%%%%%%%%%%%%%%%%%%%

\end{document}